\documentclass[12pt,preprint,notoc,nohyper]{SymmGR3} 

\usepackage{epsfig,multicol,bbm}

\newcommand\fverb{\setbox\pippobox=\hbox\bgroup\verb}
\newcommand\fverbdo{\egroup\medskip\noindent%
            \fbox{\unhbox\pippobox}\ }
\newcommand\fverbit{\egroup\item[\fbox{\unhbox\pippobox}]}
\newbox\pippobox

\title{Geometric symmetries on Lorentzian manifolds}

\author{K. Saifullah\thanks{\emph{On leave from:}
Centre for Advanced Mathematics and Physics, National University of
Sciences and Technology, Rawalpindi, Pakistan, \emph{and} Department
of Mathematics, Quaid-i-Azam University, Islamabad, Pakistan.} \\
    School of Mathematical Sciences, Queen Mary, University of London, \\ London,
UK\\
    Electronic address: \email{saifullah@qau.edu.pk}}

\preprint{}  

\abstract{Lie derivatives of various geometrical and physical
quantities define symmetries and conformal symmetries in general
relativity. Thus we obtain motions, collineations, conformal motions
and conformal collineations. These symmetries are used not only to
find new solutions of Einstein's field equations but to classify the
spaces also. Different classification schemes are presented here.
Relationships between these symmetries are discussed and
illustrating examples are presented.}

\dedicated{}

\begin{document}

\section{Introduction}
We consider a four dimensional Lorentzian manifold $M$ with metric
tensor $g_{ab}$, which is a function of
the position given in coordinates by $x^{a}$, ($a$, $b$, ...$=0$, $1$, $2$, $3$%
). If $R_{ab}$ is the Ricci tensor and $R$ is the Ricci scalar,
Einstein's Field Equations (EFEs) can be written (without
cosmological constant) as \cite{mac}

\begin{equation}
R_{ab}-\frac{1}{2}g_{ab}R=\kappa T_{ab},  \label{f11}
\end{equation}
where $T_{ab}$ is the energy-momentum tensor of the matter and
$\kappa $ is called the Einstein gravitational constant. These are
the basic equations of the theory of general relativity (GR) which
relate the geometry of the space to its matter content by virtue of
which GR becomes the theory of the dependence of the metric of a
Riemann manifold

\begin{equation}
ds^{2}=g_{ab}dx^{a}dx^{b}\,,  \label{f12}
\end{equation}
on the distribution and motion of matter.

Since $R_{ab}$ is a non-linear function of $g_{ab}$ and its first
and second derivatives, the EFEs are a system\ of 10 coupled highly
non-linear second order partial differential equations for the 10
independent functions $g_{ab}$ of four spacetime coordinates,
$x^{a}$. Solving these equations amounts to determining the 10
components of the energy-momentum tensor, as well as the 10
components of $g_{ab}$ of four variables; this makes the system
undetermined. One way to tackle this problem is by making certain
assumptions on the matter contents of the space (i.e. $T_{ab}$). On
the other hand if no particular matter distribution is assumed, then
the solutions can be obtained by imposing symmetry conditions on or
by restricting the algebraic structure of the metric, the Ricci
tensor or the Riemann tensor. Motions or Killing vectors (KVs),
which are the symmetries of the metric, and Ricci collineations
(RCs), which are the symmetries of the Ricci tensor, are two
examples of such symmetries. These symmetry properties are described
by continuous groups of motions or collineations and they lead to
conservation laws. Apart from their significance in obtaining the
exact solutions of the field equations, these symmetries provide
various invariant bases for classification of spacetimes also. The
techniques used for this purpose include application of groups of
motions and algebraic classification of the Weyl tensor  for what
are called Petrov types \cite{mac} which were understood more simply
by spinor methods developed by Penrose \cite{PenRin}. Segr\'{e}
classification \cite{mac} can also be obtained by spinors. Since the
Ricci tensor and the energy-momentum tensor are mathematically
similar, the study of RCs is important from the point of view of the
study of symmetries of matter (called matter collineations or MCs)
also, apart from their geometrical significance. Other important
symmetries in GR include homothetic motions (HMs), which are
obtained when the Lie derivative of the metric is proportional to
the metric; curvature and Weyl collineations
--- symmetries of the Riemann and Weyl tensors respectively. For
an introduction to these symmetries in GR one can see Ref. 3.

In this paper we discuss different geometric symmetries in GR and
various approaches to obtaining symmetries and classifying
spacetimes. Relationships between these symmetries are also
discussed and illustrating examples are presented.

\section{Symmetries and the Lie derivative}

`\emph{The key to symmetries is the use of the Lie derivative}'
\cite{harrison}. For each point $p$ in $M$, a vector field
$\mathbf{V}$ on $M$ determines a unique curve $\alpha _{p}(t)$ such
that $\alpha _{p}(o)=p\,$and
$\mathbf{V}$ is the tangent vector to the curve. Now, consider a mapping $%
h_{t}$ dragging each point $p$,$\,$with coordinates $x^{i}$, along
the curve $\alpha _{p}(t)$ through $p$ into the image point
$q=h_{t}(p)$, with coordinates $y^{i}(t)$. If $t\,$is very small the
map $h_{t}$ is a one-one map and induces a map $h_{t}^{*}\mathbf{T}$
of any tensor $\mathbf{T}$. The Lie derivative of $\mathbf{T}$ with
respect to $\mathbf{V}$ is defined by \cite{mac, yano}

\begin{equation}
\pounds _{\mathbf{V}}\mathbf{T=} {\lim }({t\rightarrow 0})\frac{1}{t}%
(h_{t}^{*}\mathbf{T-T)\,}.  \label{ld}
\end{equation}
Using the coordinate bases $\{\partial _{x^{i}}\}$ and $\{\partial
_{y^{i}}\}$, the Lie derivative of a vector $\mathbf{U}$ with
respect to $\mathbf{V}$ can be written as
\begin{equation}
\pounds _{\mathbf{V}}\mathbf{U=}v^{m}\frac{\partial }{\partial
x^{m}}\left(
u^{i}\frac{\partial }{\partial x^{i}}\,\right) -u^{m}\frac{\partial }{%
\partial x^{m}}\left( v^{i}\frac{\partial }{\partial x^{i}}\,\right) =\left[
\mathbf{V}\,,\mathbf{U}\right] ,
\end{equation}
where the commutator $\left[ \mathbf{V}\,,\mathbf{U}\right] $ is the
Lie bracket which is bilinear, antisymmetric and satisfies the
Jacobi identity. A linear space of smooth vector fields under the
operation of Lie bracket forms a Lie algebra. If $\left\{
\mathbf{X}_{i},i=1,...,n\right\} $ is a basis for the Lie algebra,
then we can always write
\begin{equation}
\left[ \mathbf{X}_{k},\mathbf{X}_{l}\right]
=C_{kl}^{j}\mathbf{X}_{j}
\,\,\,\,\,\,\,\,\,\,\,\,\,\,\,\,\,\,\,\,\,\,\,\,\,\,\,\,\,%
\,C_{kl}^{j}=-C_{lk}^{j}.
\end{equation}
Here $C_{kl}^{j}$ are the structure constants which completely
characterize the Lie algebra. If all the structure constants vanish
the Lie algebra is Abelian. Every Lie algebra defines a unique
simply connected Lie group and vice versa.

Lie derivatives are used in mathematical physics to express the
invariance of a tensor field under some transformation. A tensor
field $\mathbf{T}$ is invariant under a vector field $\mathbf{V}$ if the tensors $%
h_{t}^{*}\mathbf{T}$ and $\mathbf{T}$ coincide for $t$ in some
interval around $0$, i.e., the Lie derivative vanishes
\begin{equation}
\pounds _{\mathbf{V}}\mathbf{T}=0.
\end{equation}
If $\mathbf{T}$ has physical importance --- e.g., it might be the
metric tensor, or a scalar field describing a particle, or
a vector field of force --- then those special vector fields under which $%
\mathbf{T}$ is invariant will also be important.

The manifolds of interest in mathematical physics have metric
tensors. It is of interest to know when the metric is invariant with
respect to some vector field. The vector fields along which the
metric remains invariant are called Killing vector (KV) fields or
isometries. After the spacetime metric, the curvature, Ricci and
Weyl tensors are other important quantities that play a significant
role in understanding the geometric structure of spacetimes in GR.
While the isometries provide information of the symmetries inherent
in the spacetime, the Ricci Collineations (RCs), vector fields along
which the Ricci tensor is invariant under Lie transport, are
important from the physical point of view as well.

\section{Classification of spacetimes by symmetries}

Here we formally define some of the symmetries used in general
relativity  and describe various classification schemes for the
solutions of EFEs based on these symmetry methods.

\subsection{Killing vectors}

We call $\mathbf{\xi}$ a Killing vector or motion (or isometry) if
the Lie derivative of $\mathbf{g}$ with respect to $\mathbf{\xi }$
is conserved, i.e.

\begin{equation}
\pounds _{\mathbf{\xi}}\mathbf{g}=0\;.  \label{f14}
\end{equation}
This equation, called the Killing equation, in a torsion free space
and in a coordinate basis, can also be written as

\begin{equation}
g_{ab,c}\xi ^{c}+g_{ac}\xi _{,b}^{c}+g_{bc}\xi _{,a}^{c}=0\, ,
\label{f15}
\end{equation}
where ``$,$'' denotes the partial derivative. Its solutions are KVs.
The set of all solutions of Eqs. $\left( \ref{f15}%
\right) $ forms a Lie algebra and generates a Lie group of
transformations. In a four dimensional space the maximum dimension
of the Lie algebra is 10 \cite {mac}. Many explicit solutions of
EFEs have been found using Killing symmetries. KVs can be used to
derive the most general axially symmetric stationary metric
\cite{islam}. These symmetries leave all the curvature quantities
invariant and they help in describing the kinematic and dynamic
properties of spaces.

Soon after Killing's discovery of the Killing equations at the end
of the nineteenth century and Lie's classification over the complex
numbers of all Lie algebras up to dimension 3, Bianchi classified
all 3 dimensional Riemannian manifolds according to their
isometries. In this classification, which has recently been
reprinted \cite{bianchi},  Bianchi derived the full Killing vector
Lie algebra for each possible symmetry class of group actions and
solved the Killing equations to derive the metric. He also gave a
representative line element for a given symmetry type.

The attempts to classify all solutions of the EFEs on the basis of
KVs faced problems initially because of the arbitrariness of the
energy-momentum tensor. A procedure was needed which could provide a
list of all metrics according to a given isometry group and a
complete list of all isometry groups. This is an alternate approach
to solving the EFEs for given $T_{ab}$. Using this approach
Eisenhart \cite{eisen} succeeded in classifying all 2- and
3-dimensional spaces. He also developed important general theorems
concerning groups of motions. Petrov \cite{petrov} gave an invariant
classification of Riemannian spacetimes admitting groups of motions
on the basis of their detailed group structure. But his extension to
4-dimensional spaces was incomplete as admitted by him. However,
Bokhari and Qadir \cite{BQZ} classified all static spherically
symmetric spacetimes by solving the Killing equations for both
$g_{ab}$ and KVs. Obtaining the solutions for Eqs. (\ref{f15}) for
the general spherically symmetric line element in the usual
coordinates with $\nu$ and $\lambda$ as arbitrary functions

\begin{equation}
ds^{2}=e^{\nu \left(r\right) }dt^{2}-e^{\lambda \left(r \right)
}dr^{2}-r^2 d\theta ^{2}-r^2 sin^{2} \theta d\varphi^{2}
\label{sphmetric}
\end{equation}
gives a complete list of KVs for these spacetimes. The minimal
dimension of the Lie algebra is 4 and maximum 10. The minimal
symmetry is given by (writing $\frac{\partial}{\partial t}$ as
$\partial _{t}$ etc.)

\begin{equation}
\mathbf{X_1}=\partial _{t}\, , \mathbf{X_2}=sin\phi \partial
_{\theta }+cos\phi cot\theta \partial _\phi\, , \mathbf{X_3}=cos\phi
\partial _\theta-sin\phi cot\theta  \partial _\phi\, ,
\mathbf{X_4}=\partial _{\phi}\, . \label{sphmin}
\end{equation}
The maximum symmetry is admitted by the de Sitter, ant-de Sitter and
Minkowski spaces. Qadir and Ziad \cite{Thss} later provided a
complete classification of \textit{all} non-static spherically
symmetric spacetimes according to their isometries and metrics. In
this case the metric coefficients in the above metric were functions
of $t$ also. This study was extended to five-dimensional spaces
\cite{rcheu} also.

Using these methods, classifications for static spacetimes with
plane symmetry \cite{plnkv} and cylindrical symmetry \cite{QZ} were
also obtained. The stationary cylindrically symmetric fields are
hypersurface-homogeneous spacetimes and admit three Killing vectors

\begin{equation}
 \mathbf{X_1}=\partial _{t}\, , \mathbf{X_2}=\partial _{\theta }\, , \mathbf{X_3}=
\partial _{z}\, ,
\end{equation}
as the minimal symmetry which has the group $\mathbf{R} \otimes
SO(2) \otimes \mathbf{R}$. The most general cylindrically symmetric
static metric in coordinates $\left( t,\rho ,\theta ,z\right)$ can
be written as \cite{mac}

\begin{equation}
ds^{2}=e^{\nu \left( \rho \right) }dt^{2}-d\rho ^{2}-a^{2}e^{\lambda
\left( \rho \right) }d\theta ^{2}-e^{\mu \left( \rho \right) }dz^{2}
.  \label{cylmetric}
\end{equation}
These spacetimes admit groups $G_n$ of KVs, where $n=3, 4, 5, 6, 7,
10$. For the plane symmetric metric

\begin{equation}
ds^{2}=e^{\nu \left(t,x \right) }dt^{2}-e^{\lambda \left(t, x
\right) }dx^{2}-e^{\mu \left(t,x \right) } \left(dy^2+dz^2\right)
 , \label{planemetric}
\end{equation}
the minimal symmetry is given by

\begin{equation}
 \mathbf{X_1}=\partial _{y}\, , \mathbf{X_2}=
\partial _{z}\, , \mathbf{X_3}=z\partial _{y}-y\partial _{z}\, .
\end{equation}

In the static case the spacetimes admit a timelike KV,
$\mathbf{X_4}=\partial _{t}$, in addition to the KVs given above.
The dimensions of the isometry group for the associated metrics are
four, five, six, seven and ten; eight and nine are not admissible.

\subsection{Homothetic motions}

The vector $\mathbf{\xi}$ is said to be a homothetic motion (HM) if
the right hand side of Eq. (\ref{f14}) is replaced by $\phi
\mathbf{g}$, i.e.,

\begin{equation}
\pounds _{\mathbf{\xi}}\mathbf{g}=\phi \mathbf{g}\, , \label{f16}
\end{equation}
where $\phi $ is a nonzero constant.

As in the case of KVs, all the curvature quantities (except the
scalar curvature which is preserved up to a constant factor) are
invariant under a homothetic vector field. HMs give rise to
self-similar spacetimes. It is known \cite{eisen} that for the
spaces admitting $G_n$ as the maximal group of isometries, the group
$H_m$ for the HMs can be at most of the order $m=n+1$.

Hall and Steele \cite{HS} investigated the Segr\'{e} and Petrov
types of spaces that admit proper homothety groups --- HMs which are
not KVs. They classified all such gravitational fields for homothety
groups, $H_{m},\,m\geq 6,$ and gave some remarks for $m\leq 5$. Hall
\cite{hall} has also discussed the relation between homotheties and
singularities of spacetimes.

Spherically symmetric spacetimes can have homothety groups
\cite{daud} of the order $H_m, m=4,5,6,7,8,11$. For $r=11$, the only
spacetime is Minkowski.  The classification of plane symmetric
static \cite{ZiadHM} and the cylindrically symmetric static
\cite{sharif} Lorentzian manifolds according to their homotheties
and metrics is consistent with the established theorems and known
results. In the case of cylindrical symmetry some local homotheties
can be extended globally also.

\subsection{Ricci collineations}

A vector field $\mathbf{\xi}$ is an RC if the Lie derivative of a
Ricci tensor $\mathbf{R}$ with respect to $\mathbf{\xi}$ is
conserved, i.e.

\begin{equation}
\pounds _{\mathbf{\xi}}\mathbf{R}=0\; ,  \label{f17}
\end{equation}
or in component form

\begin{equation}
\xi^{c}R_{ab,c}+R_{ac} \xi_{,b}^{c}+R_{bc} \xi_{,a}^{c}=0\, .
\label{f18}
\end{equation}

There is one important point to note here that while the metric
tensor is always non-degenerate (i.e. $\det \bf{g}\neq 0$) the other
tensors like the Ricci and matter can be degenerate (i.e. the
determinant is zero) also. This gives rise to the possibility of
getting infinite dimensional algebras for the collineations.

N\'{u}\~{n}ez \emph{et al.} \cite{NPV} studied RCs of the
Robertson-Walker spacetime for which the complete solution was later
provided by Camci and Barnes \cite{cb}. Melfo \emph{et al.} \cite
{MNPV} studied RC symmetry in Godel-type spacetimes and Hall
\emph{et al.} \cite{HRV} studied RCs for various decomposable
spacetimes and discussed the relationship between RCs and matter
collineations. Camci \emph{et al.} and other authors \cite{R2451}
discussed RCs for various Bianchi type spacetimes. An interesting
study of RCs in type B warped spacetimes was done by Flores \emph{et
al.} \cite{R6}. RCs and MCs for locally rotationally symmetric
spacetimes were studied by Tsamparlis and Apostolopoulos \cite{R8}.
They studied these symmetries in Friedmann-Lemaitre universes also
\cite{R7}.

Bokhari and Qadir \cite{sp1} and later Amir \emph{et al.} \cite{sp2}
did some work on classification of static spherically symmetric
spacetimes according to their RCs . Later Qadir and Ziad \cite{sp3}
and Contreras \emph{et al.} \cite{CNP} extended this study to the
non-static case. Different aspects of RCs of this important class of
spacetimes were also studied \cite{R39}. Plane symmetry may locally
be thought of as a special case of cylindrical symmetry, therefore,
the classification for plane symmetry \cite{15} can be obtained as a
special case from the classification for cylindrical symmetry
\cite{QSZ}. Some general observations from this work are:

\begin{enumerate}
\item  For cylindrical symmetry the RC equations (\ref{f18}) are
invariant under the interchange of any two of the three coordinates
$t$, $\theta $ and $z$ (indices 0, 2 and 3).

\item Minimal symmetry for cylindrically symmetric static
spacetimes is given by $<\partial _{t}$, $\partial _{\theta }$,
$\partial _{z}>$, translations in $t$ and $z$, and rotation in
$\theta$, and has the algebra $\mathbf{R}\otimes SO(2) \otimes
\mathbf{R}$.

\item Cylindrically symmetric static spacetimes with
non-degenerate Ricci tensor admit RCs with Lie algebras of
dimensions 3, 4, 5, 6, 7 and 10. There are no 8- or 9-dimensional
Lie algebras.

\item For the degenerate Ricci tensor the RCs have infinite
dimensional Lie algebras except when $R_{11}=0$ and $R_{ii}$
($i=0,2,3$) are non-zero, the Lie algebras can be finite
dimensional.

\item For the non-degenerate Ricci tensor, if any of the $R_{00}$, $R_{22}$ or $%
R_{33}$ components are nonzero constants, the space admits
non-isometric RCs. This is an interesting and useful result for
identifying proper RCs. No such result exists in the literature for
the important class of spherically symmetric spacetimes and there is
a need to do work in this direction.
\end{enumerate}

As regards the physical significance of RCs, Davis \emph{et al.}
\cite{DK} did the pioneering work on the important role of RCs and
the related conservation laws that are admitted by particular types
of matter fields. They showed that the existence of isometries and
collineations leads to conservation laws in the form of integrals of
a dynamical system. They also considered the application of these
results to relativistic hydrodynamics and plasma physics. Oliver and
Davis \cite{OD} obtained conservation expressions for perfect fluids
using RCs. The properties of fluid spacetimes admitting RCs were
studied by Tsamparlis and Mason \cite{TM}. They have studied perfect
fluid spacetimes in detail and have also considered a variety of
imperfect fluids with cosmological constant and with anisotropic
pressure.

\subsection{Matter collineations}

If the Ricci tensor in Eqs. (\ref{f18}) is replaced by the
energy-momentum tensor then the vector field $\mathbf{\xi}$ is
called a matter collineation (MC). Since the Ricci and matter
tensors are mathematically similar, the procedure for finding MCs is
similar to that for RCs. Like KVs the maximum dimension of the Lie
algebras for RCs and MCs is also ten in four dimensional space.

Carot \emph{et al.} \cite{carot} provided one of the pioneering
studies on this symmetry. Recently, MCs for different spacetimes
have been discussed in the literature \cite{xx, R8, R7}. Various
classes of spacetimes have been classified on the basis of this
symmetry as well \cite{KS, SQ}. Solving MC equations gives rise to
various cases characterized by the constraints on the components of
the energy-momentum tensor. This includes cases of non-degenerate as
well degenerate tensors. Earlier remarks regarding RCs hold for MCs
also.

\subsection{Curvature collineations}

Replacing $\mathbf{g}$ by the Riemann curvature tensor,
$\mathbf{R}$, in component form Eq. (\ref{f14}) becomes

\begin{equation}
R_{bcd,f}^{a} \xi^{f}+R_{fcd}^{a} \xi_{,b}^{f}+R_{bfd}^{a}
\xi_{,c}^{f}+R_{bcf}^{a} \xi_{,d}^{f}-R_{bcd}^{f} \xi_{,f}^{a}=0\, ,
\label{cc}
\end{equation}
and the vector $\bf \xi$ is called a curvature collineation (CC).
This is a more complicated system as compared to those of RCs or
MCs, and is very tedious to solve completely. The maximum dimension
of a CC algebra is also ten, when it is finite. CCs for different
classes of spacetimes have been studied by Bokhari \emph{et al.}
\cite{C123}, Hall and Shabbir \cite{C47} and Shabbir \cite{C568910}.

\subsection{Weyl collineations}

Replacing $\mathbf{R}$ by the Weyl curvature tensor, $\mathbf{C,}$
in Eq. (\ref{cc}) gives Weyl collineations (WCs) \cite{ibrar}
\begin{equation}
C_{bcd,f}^{a} \xi^{f}+C_{fcd}^{a} \xi_{,b}^{f}+C_{bfd}^{a}
\xi_{,c}^{f}+C_{bcf}^{a} \xi_{,d}^{f}-C_{bcd}^{f} \xi_{,f}^{a}=0\, .
\label{wc}
\end{equation}
Very little work \cite{ibrar, W1234} has been done on this important
symmetry and there is a need to do more in this direction.

\section{Relationship between symmetries}

It is clear from their definitions that motions, affine
collineations (defined by $\pounds _{\xi }\Gamma _{bc}^{a}=0$, where
$\Gamma _{bc}^{a}$ is the Christoffel symbol) and HMs are
automatically CCs which are in turn RCs, but the converse is not
true in general. The RCs (or MCs) which are not KVs are called
proper RCs (or MCs) \cite{QSZ}. We note that RCs (and contracted
Ricci collineations) are the most general of all the symmetries.

\subsection{Isometries and other symmetries}

Isometries or KVs are the fundamental symmetries in GR. Since the
other tensors, like the Ricci and curvature tensors,  are built from
the metric tensor they must inherit its symmetries. Thus if the Lie
derivative of $\mathbf{g}$ vanishes, it must vanish for those
tensors also. Hence every KV is an RC but the converse may not be
true. We call the RCs which are not KVs ``proper RCs''. For Einstein
spaces, $\mathbf{R}\propto \mathbf{g}$, therefore, in this case the
RCs and isometries coincide. Similarly, every KV is an MC, CC or WC
but the converse is not true. As noted earlier the symmetries of the
metric always have finite dimensional Lie algebras, which is not the
case with those of the other tensors, in general.

\subsection{Ricci and matter collineations}

As mentioned earlier, if the components of the Ricci tensor,
$R_{ab}$, in Eq. (\ref{f18}) are replaced by those of the
energy-momentum tensor, we get MCs. But this does not mean that for
a given space RCs and MCs will be identical. In some spaces RCs are
greater than MCs, while in others MCs are more than RCs, which shows
that neither of the sets contains the other, in general \cite{KS}.
In some cases they may coincide as well. Let us consider the
following metric for comparison of symmetries

\[
ds^{2}=\left( x/x_{0}\right) ^{2a}dt^{2}-dx^{2}-\left(
x/x_{0}\right) ^{2}\left( dy^{2}+dz^{2}\,\right) \, ,
\]
where $a$ and $x_{0}$ are constants and $a\neq 0,1,-1$. For this
metric $R_{ab}$ are given by
\begin{equation}
\left.
\begin{array}{l}
R_{00}=a\left( 1+a\right) x^{2a-2}/x_{0}^{2a}\, , \\
R_{11}=-\left( -a+a^{2}\right) /x^{2}\, , \\
R_{22}=-\left( 1+a\right) /x_{0}^{2}=R_{33} .
\end{array}
\right.
\end{equation}
The energy-momentum tensor for this metric can be written as

\begin{equation} \left.
\begin{array}{l}
T_{00}=-x^{2a-2}/x_{0}^{2a}\, , \\
T_{11}=\left( 2a+1\right) /x^{2}\, ,\\
T_{22}=a^{2}/x_{0}^{2}=T_{33}\, .
\end{array}
\right.
\end{equation}
One may note that the energy density is negative and cannot be made
positive by introducing a cosmological constant. The dimension of
the KV algebra for this metric is four, while that of RCs and MCs is
six, and it is spanned by

\begin{equation}
\begin{array}{l}
\mathbf{X}_{1}=\partial _{t}\,, \\
\mathbf{X}_{2}=\partial _{y}\,, \\
\mathbf{X}_{3}=\partial _{z}\,, \\
\mathbf{X}_{4}=z\partial _{y}-y\partial _{z}\,\,, \\
\mathbf{X}_{5}=\left( \frac{1}{T_{00}}-\frac{1}{4}t^{2}\right)
\partial _{t}+\frac{1}{\sqrt{T_{11}}}t\partial _{x}\,, \\
\mathbf{X}_{6}=-\frac{1}{2}t\partial _{t}+\frac{1}{\sqrt{T_{11}}}
\partial _{x}\,.
\end{array}
\end{equation}

The algebra is given by

\[
\begin{tabular}{lll}
$\left[ \mathbf{X}_{1},\mathbf{X}_{5}\right] =\mathbf{X}_{6} \,,$ &
$\left[ \mathbf{X}_{1},\mathbf{X}_{6}\right]
=-\frac{1}{2}\mathbf{X}_{1}\,,$
& $\left[ \mathbf{X}_{2},\mathbf{X}_{4}\right] =-\mathbf{X}_{3} \,,$ \\
$\left[ \mathbf{X}_{3},\mathbf{X}_{4}\right] =\mathbf{X}_{2} \,,$ &
$\left[ \mathbf{X}_{5},\mathbf{X}_{6}\right]
=\frac{1}{2}\mathbf{X}_{5}\,,$ &
$\left[\mathbf{X}_{i},\mathbf{X}_{j}\right] =0 \,,$ {otherwise.}
\end{tabular}
\]
This is a direct sum of the two subalgebras $\left\{ \mathbf{X}_{2},
\mathbf{X}_{3},\mathbf{X}_{4}\right\} $ and $\left\{
\mathbf{X}_{1},\mathbf{X}_{5},\mathbf{X}_{6}\right\}$.

\subsection{Curvature and Weyl collineations}

Though the Weyl and Riemann curvature tensors have similar forms,
the local symmetries of the curvature tensor (i.e. CCs) and the Weyl
tensor (WCs) are different. One can write a metric with a finite
dimensional Lie algebra of Weyl symmetries that properly contains
the Lie algebra of curvature symmetries \cite{ibrar}. However, there
is no example known for the converse case.

Using the tensorial symmetries of the curvature or Weyl tensor, in
component form they can be written as 6$\times 6$ matrices. Thus
they are degenerate if they are of rank 5 or less. It is possible
that one of them is degenerate and the other non-degenerate. An
example is the de-Sitter (or anti de-Sitter) spacetime for which the
Lie algebra of CCs is finite dimensional while that of WCs is
infinite dimensional such that every vector field is a WC. The
question arises whether there are cases in which the set of WCs is
properly contained in the set of CCs when both have finite
dimensional Lie algebras. So far, no such example exists in the
literature; there is neither a proof that this cannot be the case.
For the comparison of WCs and CCs we consider the metric
\cite{ibrar}

\begin{equation}
ds^{2}=dt^{2}-d\rho^{2}-(\rho/a)^2 d\theta ^{2}-(\rho/a)^2 dz^{2}.
\end{equation}
In this case the stress-energy tensor is given by

\[
T_{00}=-\frac{1}{\kappa \rho^{2}}=-T_{11}\, , T_{22}=0=T_{33}\, ,
\]
so that it is not a physically realistic spacetime. The non-zero
component of the curvature tensor is
\[
R_{323}^{2}=-\frac{1}{a^{2}}
\]
and those of the Weyl tensor are
\begin{eqnarray*}
C_{101}^{0} &=&-\frac{1}{3\rho^{2}}\, ,
C_{202}^{0}=\frac{1}{6}=C_{212}^{1}\, ,
\\
C_{303}^{0} &=&\frac{1}{6a^{2}}=C_{313}^{1}\, ,
C_{323}^{2}=-\frac{1}{3a^{2}}\, .
\end{eqnarray*}
Here the Ricci tensor is degenerate. This space has one extra KV
\[\mathbf{X}_{4}=-\frac{z}{a}\frac{\partial }{\partial \theta }+a\theta
\frac{\partial }{\partial z}\, ,
\]
one proper HM
\[\mathbf{X}_{5}=t\frac{\partial }{\partial t}+\rho\frac{\partial
}{\partial \rho}\, ,\] and one additional WC
\[
\mathbf{X}_{6}=\frac{1}{2}(t^{2}+\rho^{2})\frac{\partial }{\partial
t}+t \rho \frac{\partial }{\partial \rho}\, .
\]
There are infinitely many CCs and RCs. Thus we see that in this case
the set of KVs is contained in that of HMs which are a subset of WCs
which are contained in CCs.

\section{Summary and conclusion}

Motions preserve distances. Conformal motions preserve angles
between two directions at a point and map null geodesics into null
geodesics. HMs scale all distances by the same constant factor,
therefore, they lead to self-similar spacetimes. HMs also preserve
the null geodesic affine parameters. Projective collineations

\[
\pounds _{\xi }\Pi _{bc}^{a}=0\,,
\]
where the projective connection $\Pi _{bc}^{a}$ is given by
\[
\Pi _{bc}^{a}=\Gamma _{bc}^{a}-\frac{1}{n+1}\left( \delta
_{b}^{a}\Gamma _{dc}^{d}+\delta _{c}^{a}\Gamma _{db}^{d}\right) ,
\]
map geodesics into geodesics and affine collineations preserve, in
addition, the affine parameters on geodesics \cite{mac}.

New solutions of EFEs can be constructed by using symmetries. When
the RC equations, for example, are solved to obtain the RC vectors,
we also obtain some (differential) constraints on the Ricci tensor
of the space. Solving these constraints gives the metric \cite{QSZ}.
Another application of symmetries is that the solutions can be
classified on their basis. We express symmetries in terms of Lie
algebras which can be of finite or infinite dimensions.

Motions or KVs constitute the basic symmetry in GR. They form a
subset of HMs which are contained in CCs. All these symmetries are
subsets of RCs. A few symmetries have been described in some detail.
The physical significance of KVs and HMs and their role in
conservation laws is well understood. Similarly, WCs as the
symmetries of the gravitational field and MCs of the matter field
are important. However, the role of other symmetries from a physical
point of view  is yet to be understood \cite{symm}. Mutual
relationship between some of these motions and collineations has
been discussed. In particular, the inclusion relationships between
RCs and MCs, and CCs and WCs have been explored. There is room for
further research in this direction and some open problems have been
mentioned.

We may mention here that besides these symmetries which have been
described in this paper, there are other symmetries discussed in the
literature \cite{katzin, symm} on GR also.

\acknowledgments

The author is thankful to Professor Asghar Qadir for reading the
manuscript and suggesting some improvements. A research grant from
the Higher Education Commission of Pakistan is acknowledged.

\end{document}